\documentclass[a4paper,11pt]{article}
\usepackage{jcappub}
\usepackage{hyperref}

\title{Low-$\ell$ power suppression in punctuated inflation }
\author[a]{Mussadiq H. Qureshi,}
\author[a,1]{Asif Iqbal \note{Corresponding author.},}
\author[a,b]{Manzoor A. Malik}
\author[b]{and Tarun Souradeep} 
\affiliation[a]{Department of Physics, University of Kashmir, Hazratbal, \\
Srinagar, Jammu and Kashmir 190006, India  \\}
\affiliation[b]{Inter-University Center for Astronomy and Astrophysics, Post Bag 4, Ganeshkhind, \\
Pune 411007, India}
\emailAdd{mussadiqqureshi111@gmail.com}
\emailAdd{asifiqbal@kashmiruniversity.net} 
\emailAdd{mmalik@kashmiruniversity.ac.in}  
\emailAdd{tarun@iucaa.in}
\abstract{ Motivated by Planck confirmation of an anomalously low value of the CMB temperature fluctuations up to
multipole $\ell<40$, we in this paper try to explain such feature by investigating case of punctuated inflation scenario. This form of inflation potential is inspired by Minimal Super-symmetric Standard Model (MSSM) wherein suppression of curvature perturbation power at large scales
is produced  by introducing period of fast-roll phase of the inflation sandwiched between two stages of slow-roll phase. 
   
We apply Markov Chain Monte Carlo analysis to determine posterior distribution and the best fit values  of the model parameters  using recent WMAP9 and Planck
data. We show that  WMAP9 and Planck results are consistent with each other and that with Planck data we obtain tighter constraints for punctuated inflation parameters.
We find that punctuated inflation leads to better fit in CMB data compared to simple power law model. 
The improvement in the fit to the WMAP9 data is $\Delta \chi^2\sim 3.6$ and for Planck the improvement is $\Delta \chi^2\sim 5.4$.  We find 
that $AIC$ does not discriminate between punctuated inflation and simple power law model for WMAP9 data. However, for Planck data we find that punctuated inflation is moderately preferred  over a simple power law model.}
\begin{document}
\maketitle
\section{Introduction}
Anisotropy in the cosmic microwave background (CMB) is one of the most robust and informative tools in cosmology. Measurements of the CMB anisotropies
by COBE, WMAP9 \cite{Smoot1992,Hinshaw2013} and Planck \cite{Ade2013,Ade2015} are in striking agreement with $\Lambda$CDM cosmology.
These results are consistent with an adiabatic and nearly scale invariant spectrum of primordial perturbations, as predicted by the simple inflationary models \cite{Guth1981,Liddle1994,Liddle2000}.
The inflationary model(s) provide us potential solutions to most of the outstanding problems of standard Big Bang. They
successfully explain the flatness problem, horizon problem and the monopole problem \cite{Guth1981,Liddle1994,1993PhR...231....1L}. A key feature of these models is that during inflationary epoch, in the very early universe, primordial quantum
fluctuations \cite{Hawking1982,Starobinsky1982,Guth1982} of some scalar fields are produced which eventually transform into large scale structure of the universe under the gravitational instability paradigm. The observed
statistical distribution of temperature fluctuations in the cosmic microwave background is believed to be sourced by these quantum fluctuations.

Although, power law (PL) model with almost scale-invariant, adiabatic and Gaussian primordial spectrum ``$\mathcal{P}(k)$'' ($k$ being the co-moving wavenumber)  has emerged as the most successful model from the recent observations, there exist several anomalies in CMB data which could have far reaching consequences. 
Some of these anomalies are low CMB power at large angular scales \cite{Bond1998,Sinha2006,Cicoli2014,Iqbal2015}, 
departure from either Gaussianity or statistical isotropy \cite{2003ApJ...597L...5H,Seery2007,Aich2010,Ade2013b,Aluri2015,Byrnes2015},  lack of correlations on large scales \cite{Copi2007,Copi2013,Gruppus2014}, hemispheric asymmetry and the cold spot \cite{Eriksen2004,2004MNRAS.354..641H,Cembranos2008,2014ApJ...784L..42A,Assadullahi2015,Ade2015c}, 
the so-called Axis of Evil \cite{Land2005,Frommert2010,Rassat2014} and oscillatory features in the primordial power spectrum \cite{Pahud2009,Flauger2010,Kobayashi2011,2012PhRvD..85b3531A,Aich2013,Peiris2013,Benetti2013,2014A&A...571A..22P,Ade2015b}. 
Many inconsistencies in angular power spectrum have also been found between WMAP9 and Planck  data \cite{2014PhRvD..89d3004H,2015ApJ...801....9L,2012JCAP...08..002S}.  Recently, Addison et al. \cite{Addison2015} found internal inconsistencies between high ($\ell>1000$) and low ($\ell<1000$) multipoles in the Planck data.
They suggest that tension in the Planck high $\ell$ spectrum is either due to an unlikely statistical fluctuation or unaccounted systematics in data.
Similarly, Hazra \& Shafieloo \cite{2014JCAP...01..043H} found that Planck data is consistent with the concordance $\Lambda$CDM  model only at 2-3$\sigma$ confidence level with an indication of lack of power at both high and low $\ell$'s with respect to concordance model. 
The study of these anomalies and their statistical significance is of utmost importance for cosmology as  these results could test the $\Lambda$CDM model.

Lack of power at low CMB multipoles can be attributed to spatial curvature \cite{2003MNRAS.343L..95E}, non-trivial topology \cite{2003Natur.425..593L},
violation of statistical anisotropies \cite{2003ApJ...597L...5H}, hemispherical anisotropy and non-gaussianity \cite{McDonald2014b,McDonald2014}, bouncing inflation \cite{2004PhRvD..69j3520P,Liu2013}, dark energy during the inflation \cite{2004PhRvD..70h3003G},  primordial micro black holes \cite{Scardigli2011}, 
loop quantum cosmology \cite{2014CQGra..31e3001B},  string theory \cite{2012JCAP...05..012D,2015MPLA...3050137K,2015PhRvD..92l1303C}, non-attractor initial conditions \cite{2016JCAP...10..017C} and so on. Besides, there exist many other classes of inflationary models which give rise to additional features in primordial power spectra compared to simple power law model. 
Exploring different inflationary models is still an interesting problem and could help us 
in explaining many anomalies.  In our earlier work, we carried a detailed study of the inflationary models which could produce cut off in primordial power spectrum at low $k$ so as to reproduce necessary suppression in the 
CMB power spectrum up to multipoles $\ell \leq40$ observed in the Planck data. We found marginal preference of the cut off parameters that describe the power suppression at low multipoles \cite{Iqbal2015}.
However, in all these models,  we have to assume  a specific pre-inflationary era (like kinetic or radiation dominated regime) and/or some special initial conditions to achieve a near scale invariant power spectrum along with a cut off at large scales (i.e low $k$).
In this paper, we will consider punctuated inflationary scenario (PI model) which allows a brief period of fast roll sandwiched between two stages of slow roll inflation.
Such  scenario can be  encountered in the Minimal Supersymmetric Standard Model (MSSM) wherein the inflationary potentials contain a point of inflection \cite{2006PhRvL..97s1304A,2007JCAP...06..019A,2007JCAP...01..015B,2011PhRvD..84j1301A,2013JCAP...07..041C}.
The break in the slow roll will induce power suppression followed by rise  in the primordial power spectrum corresponding to the scales that leave the Hubble radius just before 
the transition to the fast roll \cite{2001PhRvD..64b3512L,2007JCAP...10..003J,2009JCAP...01..009J}. The form of inflationary potential that we consider in this work has been previously investigated by Jain et al. \cite{2009JCAP...01..009J,2010PhRvD..82b3509J} using  WMAP5 year data and this work can 
be regarded as a progression of their work. They showed that brief period of fast roll induces step like feature in the primordial spectrum corresponding to the Hubble scales leading to the suppression in the CMB power at low multipoles. Such scenario is advantageous in that we don't have to consider any special pre-inflationary era or impose any special initial condition. Given the new Planck data and improved WMAP9 data, we aim to  tighten the constraints on the parameter space with respect to previous work.
Indeed, if the Planck and WMAP results are not consistent with each other then this would clearly imply the rejection of such model.  
Moreover, we also use Akaike information criterion ($AIC$) \cite{Akaike1974_aic} to compare punctuated inflation model  with PL model.

The plan of this paper is as follows: In section 2, we discuss punctuated inflationary model which can produce cut off primordial power spectrum at large scales.
We then discuss CMB data set and methodology used in  our analysis in section 3. In section 4, we give parameter estimates and explore Akaike information criterion for comparing
punctuated inflationary model with power law model. Finally, we conclude our work in section 5. Throughout this work we will assume $\hbar=c=8\pi G=1$.

\section{Punctuated inflation}
\begin{figure*}
\centering
\includegraphics[width=11cm,height=11cm,keepaspectratio]{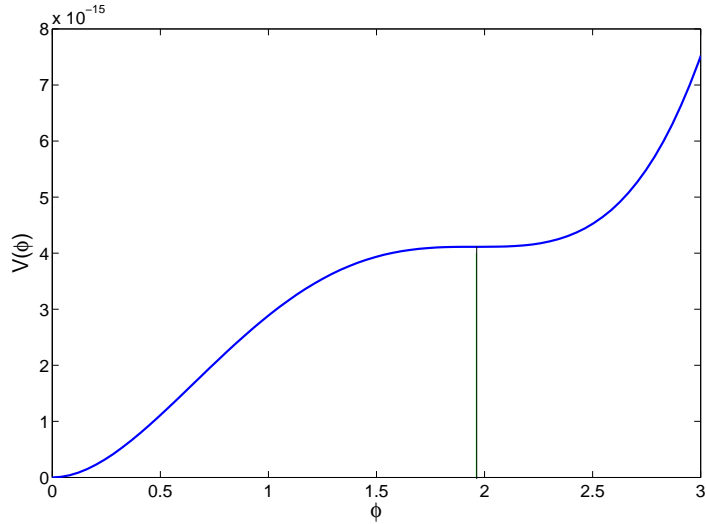}
\caption{Form of the inflationary potential for parameters: $m=1.1323\times10^{-7}$ and $\lambda=3.3299\times10^{-15}$ (or $\phi_0=1.9622$). 
The vertical line represents the inflection point.}
\label{bestfit_potential}
\end{figure*}

The deviations from the scale invariance of the power spectrum can be produced by introducing two or more periods of fast roll phase. 
Such cases arise in double inflationary scenarios where one uses two scalar fields \cite{1995PhLB..356..196P,1997NuPhB.503..405A,2003PhRvD..67h3516T,2016PhRvD..93b5001S}.
Within single field inflationary scenarios, such departure from the slow roll phase are usually produced by introducing step or sudden change in slope of the inflationary field \cite{2001PhRvD..64l3514A,2005JCAP...07..015G,2014JCAP...09..039M,2016JCAP...06..027C}.
However, there also exist many smooth and well behaved potentials that can incorporate fast roll phase \cite{1990NuPhB.335..197H,2001PhRvD..64b3512L,2007JCAP...10..003J}.
The extent of deviations would depend on the nature and duration of the fast roll which are essentially controlled by the model parameters.
The potential that we consider in this work to achieve a phase of fast roll is motivated by MSSM, an  extension of the Standard Model (SM) which has many cosmological consequences \cite{2006PhRvL..97s1304A,2007JCAP...06..019A,2007JCAP...01..015B,2011PhRvD..84j1301A,2013JCAP...07..041C}. 
In such  scenarios, inflation carries the standard model (SM) charges and eventually decays into the SM baryons and the cold dark matter. 
We will consider following form of inflationary potential which has been earlier studied by  Jain et al. \cite{2009JCAP...01..009J,2010PhRvD..82b3509J} 
\begin{equation}
V(\phi) = \left(\frac{m^2}{2}\right)\,\phi^2  - \left(\frac{\sqrt{2\,\lambda\,(n-1)}\,m}{n}\right)\, \phi^n + \left(\frac{\lambda}{4}\right)\,\phi^{2(n-1)},
\label{eq:mssm-p}
\end{equation}
where we consider the case $n=3$ and the coefficient of the $\phi^n$ term has been  chosen in such a way that the potential has a point of inflection (point where both $V_{\phi}\equiv 
(dV/d\phi)$ and $V_{\phi\phi}\equiv (d^{2}V/d\phi^{2})$ vanish) at 
$\phi = \phi_0$ given by
\begin{equation}
\phi_{0} = \left(\frac{2\,m^2}{(n-1)\,\lambda}\right)^{\frac{1}{2\,(n-2)}}.
\end{equation}
This form of the inflationary potential is shown in figure~\ref{bestfit_potential} using best fit parameters.  It is important to note that in the usual Minimal Super-symmetric Standard Models, the point of inflection is located at the  sub-Planckian values (i.e. $\phi_0<<1$). However, for our case the inflationary 
potential has an inflection at  $\phi_0>>1$, if we have to induce a period of the fast roll followed by a second phase of slow roll. This is helpful   in our case in the sense that the inflationary phase becomes 
independent of the initial value of the $\phi_i$ and $\dot \phi_i$ while in case of MSSM they have to be severely fine tuned; $\phi_i\sim\phi_0$ and $\dot \phi_i\sim0$ \cite{2006PhRvL..97s1304A}. 
It has been shown by Jain et al. \cite{2009JCAP...01..009J} that if we start with any initial value  $\phi_i>> \phi_0$ then the  subsequent dynamics of the punctuated field 
approach the attractor and the  attractor trajectory exhibit two regimes of slow roll inflation sandwiching a period of fast roll.
It is worth mentioning that even though we consider the  parameters of the inflationary potential that are different from the MSSM case, it may be possible to realize such scenarios beyond standard model (see \cite{2009JCAP...01..009J} for more details).

It is often found that if we consider single field inflationary models along with the slow roll approximation then the amplitude of the scalar perturbations freezes after the modes leave the Hubble radius which can be expressed in terms of inflationary potential 
as follows \cite{Liddle2000,1992PhR...215..203M}
\begin{equation}
{\cal P}_{S}(k)  \simeq \left(\frac{1}{12\,\pi^2}\right)\,  \left(\frac{V^3}{V_{\phi}^2}\right).
\end{equation} 
However, if we also consider an intermediate period of fast roll, it is found that the  super-Hubble amplitude of the modes that exit
the Hubble scale just before start of fast roll are enhanced compared to their value at Hubble exit. For modes that exit the Hubble scale during fast roll, the amplitudes are actually suppressed \cite{2001PhRvD..64b3512L,2008JCAP...06..024S,2007JCAP...10..003J,2016JCAP...03..028C,Cicoli2014}.
Further, the  modes that leave well before and after fast roll remain unaffected. 

\section{Solving primordial power spectrum}
In this section, we will discuss equations of motion governing the evolution of scalar (curvature) perturbations
and  tensor perturbations. We use the formalism similar to Jain et al.  \cite{2009JCAP...01..009J}, Aich et al. \cite{Aich2013} and Hazra et al. \cite{2013JCAP...05..026H} where they evolve the perturbations using the exact numerical approach using e-folds as the independent variable.
\subsection{Background equations}
The general action of scalar field in the curved space-time can be described by \cite{Weinberg1972,Padmanabhan2010}
\begin{equation}
 S\,=-\, \int d^4\,x\,\sqrt{|g|}\,\left[\mathcal{L}_g+\mathcal{L}_\phi\,\right],
\end{equation}
where 
\begin{equation}
 \mathcal{L}_g=\frac{1}{2}R, \quad \mathcal{L}_\phi =\frac{1}{2}g^{\mu \nu}\partial_{\mu}\phi \partial_{\nu}\phi - V(\phi),
\end{equation}
are the Lagrangians for the gravitation and scalar fields respectively,  $V(\phi)$ is the potential energy of the scalar field and $R$  is the curvature scalar. 
Variation of the action with respect to $\phi$ leads to the Klein-Gordon equation \cite{Padmanabhan2010}
\begin{equation}
 \frac{1}{\sqrt{|g|}}\partial_{\mu}\left[\sqrt{|g|}g^{\mu \nu}\partial_{\mu} \phi\right] + V_{\phi}=0.
\end{equation}
 For the spatially flat FRW metric where $g_{\mu \nu}=\textrm{diag}(1, -a^{2}, -a^{2}, -a^{2}$) and $\sqrt{|g|}=a^3$, $a$ being the scale factor, the above equation reduces to
\cite{1993PhR...231....1L,Liddle2000}
\begin{equation}
\ddot \phi+3H\dot \phi-a^{-2}\nabla^2\phi+V_{\phi}=0,
\label{background}
\end{equation}
where the overdots represent the differentiation with respect to cosmic time.
The second term provides friction term for the scalar field which is proportional to the Hubble constant (expansion rate) $H = \dot {a}/a$ and third term is approximately equal to zero as the inflation would rapidly smooth out spatial variations in the universe.
Similarly, variation of the action with respect to $g^{\mu \nu}$ gives us  the Einstein tensor $G_{\mu \nu}$
\begin{equation}
 G_{\mu \nu}=T_{\mu \nu},
\end{equation}
where $T_{\mu \nu}$ is energy-momentum tensor of the scalar field given by,
\begin{equation}
T_{\mu \nu}=\partial_\mu \phi \partial_\nu \phi-g_{\mu \nu} \mathcal{L}_\phi.
\end{equation}
For  spatially flat FRW metric $T_{\mu \nu}$ reduces to \cite{Padmanabhan2010}
\begin{equation}
T_{00}=\rho_{\phi} = \frac{1}{2} {\dot \phi}^2 + V (\phi), \quad  T_{ii}=p_{\phi} = \frac{1}{2} {\dot \phi}^2 - V (\phi),
\end{equation}
where $\rho_{\phi}$ and $p_{\phi}$ are the energy density and pressure density of the scalar field.
The combination of $G_{00}$ and $G_{ii}$ provides  a useful equation,
\begin{equation}
 \dot H=\frac{1}{2} \dot \phi^2, \quad 3H^2 = \frac{1}{2} {\dot \phi}^2 + V (\phi),
\label{Heqn} 
\end{equation}
where the second equation is called the Friedmann equation. 
For inflation to happen, the scalar field is assumed to vary slowly with time (i.e $\dot \phi/2<<V$) which leads to the  exponentially growing scale factor in the Friedmann equation. 
Slow-roll inflation is characterized by two parameters $ \epsilon<<1$ and $\eta<<1$ which are defined as \cite{Liddle1994}
\begin{equation}
\epsilon =  -\frac{\dot H}{H^2} =\frac{1}{2} \left(\frac{V_{\phi}}{V(\phi)} \right)^2, \quad \eta =\epsilon+\delta = \left ( \frac{V_{\phi\phi}}{V(\phi)} \right),
\end{equation}
where $\delta = - \ddot \phi/H{\dot \phi}$.
\subsection{Perturbation equations}
In spatially flat Friedmann universe, the Fourier modes of the curvature perturbation $\mathcal{R}$ and the tensor perturbation $h$ are described by the following equations \cite{1992PhR...215..203M}
\begin{equation}
 \mathcal{R}_k''+2\frac{z'}{z}\mathcal{R}_k'+k^2\mathcal{R}_k=0,  \quad  {h}_k''+2\frac{a'}{a}{h}_k'+k^2h_k=0,
\label{perturbation}
 \end{equation}
where $z=a\dot \phi/H$ and the primes represent differentiation with respect to the conformal time. 
It is advantageous to consider the gauge invariant Mukhanov variables $u$ and $v$,
\begin{equation}
 u=z\mathcal{R}, \quad  v=ah.
\end{equation}
The Fourier components $u_k$ and $v_k$ obey the following equations of motion
\begin{equation}
 u_k''+\left(k^2-\frac{z''}{z}\right)u_k=0, \quad  v_k''+\left(k^2-\frac{a''}{a}\right)v_k=0,
\end{equation}
We can then define the primordial power spectrum of curvature perturbations $\mathcal{P}_s(k)$ and tensor perturbations $\mathcal{P}_t(k)$  by the two-point 
correlation functions as \cite{Durrer1998}
\begin{equation}
 <\mathcal{R}_{k_1}\mathcal{R}^*_{k_2}>=\frac{2\pi^2}{k^3}\mathcal{P}_s(k)\delta^{(3)}(k_1-k_2), \quad  <{h}_{k_1}{h}^*_{k_2}>=\frac{2\pi^2}{k^3}\mathcal{P}_t(k)\delta^{(3)}(k_1-k_2).
\end{equation}
Assuming Gaussianity and adiabaticity, $\mathcal{P}_s(k)$ and $\mathcal{P}_t(k)$ encode all the  information for a complete statistical description of the fluctuations and are  related to 
$u_k$ and $v_k$ via
\begin{equation}
\mathcal{P}_s(k)=\frac{k^3}{2\pi^2}\left|\frac{u_k}{z}\right|^2, \quad \mathcal{P}_t(k)=2\frac{k^3}{2\pi^2}\left|\frac{v_k}{a}\right|^2.
\end{equation}
Since, tensor-to-scalar ratio ``$r$'' remains smaller than $10^{-4}$ over whole cosmological scale for our model, therefore, we shall only consider scalar perturbation
to minimize the computation time. 
\subsection{Initial conditions}
In order to evaluate scalar power spectrum for different modes of $k$,  we need to solve together background equation (equation~\ref{background}) and perturbation equation (equation~\ref{perturbation}). Since both the equations are 
second order differential equations, we need 4 initial conditions - $\phi_i$, $\dot \phi_i$, $\mathcal{R}_i$ and $\mathcal{R'}_i$. 
We assume $\phi_i$ to be equal to $12$ and $\dot \phi_i$ is evaluated using,
\begin{equation}
 3H\dot \phi_i=-V(\phi_i).
\end{equation}
 The initial conditions for $\mathcal{R}_i$ and $\mathcal{R'}_i$  can be obtained 
by imposing Bunch-Davies vacuum of de Sitter space when the perturbations are well inside the horizon,
\begin{equation}
\mathcal{R}_i =\frac{1}{\sqrt{2k}z_i}\exp(-i k\eta),
\end{equation}
where conformal time $\eta$ is an irrelevant phase. In simple inflationary models, one usually imposes the initial conditions on the modes when $k/(aH)\approx100$.  Similarly, to compute the spectrum 
we need  to integrate until the mode is far outside the horizon (i.e super-Hubble scales), typically, $k/aH\approx10^{-5}$ \cite{2013JCAP...05..026H,2009JCAP...01..009J}.
\section{Methodology}
\subsection{CMB analysis}
The initial power spectrum $ \mathcal{P}(k)$  is related to the angular power spectrum $C_{\ell}$ through,
\begin{equation}
C_\ell^{{XX'}}  \propto \int {\rm d}{\rm ln} k \, \mathcal{P}(k) \, T_\ell^{X} (k) \, T_\ell^{X'} (k),
\end{equation}
where $T_\ell^X(k)$ is the transfer function  with $X$ representing the CMB temperature or polarization. 
The measured angular power spectrum $C_l$ is a robust cosmological probe in constraining cosmological models, the position and amplitude of the peaks 
being very sensitive to important cosmological parameters.

We have used modified version of CAMB \cite{2000ApJ...538..473L,2002PhRvD..66b3531L} which is based on line of sight integration approach given in \cite{1996ApJ...469..437S}  to calculate the angular power spectra of the CMB anisotropies for different inflationary models discussed above.
For exploring the likelihood function $\mathcal{L}(\mathbf{D} |\boldsymbol{\theta},\mathcal{M})$ (or the posterior distribution) of the parameters, 
we use Markov chain Monte Carlo (MCMC) methods. We have obtained the best fit values for the parameters using modified version of publicly available CosmoMC code \cite{2002PhRvD..66j3511L} with a convergence diagnostics done through the Gelman and Rubins statistics. 
We have used PL model along with the PI derived scalar power spectrum for the cosmological estimates.
We have conducted the analysis with two different data sets: WMAP9 and Planck. For WMAP9 data set, we have used full temperature and polarization 9 year data and likelihood  code provided by the WMAP9 team. 
For Planck data set,  we have used low $\ell$ TEB likelihood  ($2 \leq \ell\leq 29$) and high $\ell $ nuisance-marginalized  Plik lite likelihood ($30 \leq \ell\leq 2500$) to compute joint likelihood for TT, EE, BB, TE.

\subsection{Model parameters and priors}
The cosmological parameterization has been carried out by using the six basic parameters (baryon density ($\Omega_bh^2$), cold dark matter density ($\Omega_{c}h^2$), Thomson scattering optical depth due 
to re-ionization ($\tau$),  angular size of horizon ($\theta$), scalar spectral index ($n_s$) and scalar amplitude 
($\ln10^{10}A_s$)) along with the three parameters which describe punctuated inflation ($\phi_0$, $\ln[10^{10}m^2]$) and scale factor of the universe at start of inflation ($a_0$). 
The remaining cosmological parameters are kept constant. We have fixed the sum
of physical masses of standard neutrinos ``$\nu" = 0.06$ eV, effective number of neutrinos ``Neff"$=3.046$, Helium mass fraction ``YHe"$=0.24$ and the width of re-ionization $=0.5$. 
The results of the Bayesian parameter estimation may depend on the range of priors used. Table~\ref{priorranges} shows the prior ranges of all parameters used in this work. Moreover, we have used flat prior distributions for all cosmological parameters.
\begin{table} 
\centering
\begin{tabular}{|l|c|c|c|}
\hline
Parameter Name& Symbol & Lower limit & Upper limit\\ 
\hline 
Baryon density & $\Omega_b{h}^2$ & 0.005 &0.1 \\ 
Cold dark matter density & $\Omega_{c}{h}^2$ & 0.001 &0.99 \\ 
Angular size of acoustic horizon& $\theta$ & 0.5 &10.0 \\
Optical depth & $\tau$ & 0.01 &0.8  \\ 
Scalar spectral index & $n_s$  &  0.5 &1.5 \\ 
Scalar amplitude & $\ln[10^{10} A_s]$ & 2.7 &4.0\\
Initial scale factor & $a_0$ & 0.0001 &0.1 \\
Inflection point & $\phi_0$ & 1.950 & 1.970\\
Inflationary mass & $\ln[10^{10}m^2]$ & -12 &-8 \\
\hline
\end{tabular}
\caption{Uniform prior used in parameter estimation.}
\label{priorranges}
\end{table}
\begin{table*}
\noindent\resizebox{\linewidth}{!}{
\centering
\begin{tabular}{|c|c|ccc|ccc|}  \hline
\multicolumn{2}{|c|}{} & \multicolumn{3}{c|}{WMAP9} &\multicolumn{3}{c|}{Planck}             \\  \hline
Model   & Parameter          &  Best Fit   & 68\% Limit &$\chi^2 =-2\log {\mathcal L}$                 & Best Fit   &  68 \% Limit  &$\chi^2 =-2\log {\mathcal L}$  \\          \hline 

PL     &  $\Omega_b{h}^2$                 &$ 0.0223 $  &$ 0.0223^{+0.0005}_{-0.0005} $   & $7557.828$  & $0.0211 $      &$ 0.0211^{+0.0001}_{-0.0001} $     &  $11629.944$ \\          [1ex]
       &  $\Omega_{c} h^2 $               &$ 0.1141 $  &$ 0.1138^{+0.0046}_{-0.0044} $   &             & $0.1239 $      &$ 0.1241^{+0.0015}_{-0.0015} $     &     \\                   [1ex] 
       &  $\Omega_\Lambda$                &$ 0.7110 $  &$ 0.7124^{+0.0288}_{-0.0234} $   &             & $0.6521 $      &$ 0.6506^{+0.0107}_{-0.0105} $     &      \\                  [1ex]
       &  $\ln[10^{10} A_s]$              &$ 3.0869 $  &$ 3.0878^{+0.0277}_{-0.0315} $   &             & $3.0351 $      &$ 3.0421^{+0.0382}_{-0.0393} $     &       \\                 [1ex]  
       &  $n_s$                           &$ 0.9670 $  &$ 0.9683^{+0.0126}_{-0.0127} $   &             & $0.9483 $      &$ 0.9474^{+0.0046}_{-0.045} $      &       \\                 [1ex]
       &  $\tau$                          &$ 0.0861 $  &$ 0.0875^{+0.0127}_{-0.0147} $   &             & $0.0500 $      &$ 0.0530^{+0.0193}_{-0.0196} $     &        \\                [1ex]   
       &  $100\theta$                     &$ 1.0393 $  &$ 1.0395^{+0.0057}_{-0.0073} $   &             & $1.0400 $      &$ 1.0400^{+0.0003}_{-0.0003} $     &          \\              [1ex]
       &  $z_{re}$                        &$ 10.551 $  &$ 10.617^{+1.112}_{-1.101}   $   &             & $7.5607 $      &$ 7.7661^{+2.3591}_{-1.936}  $     &           \\             [1ex]      
       &    $H_0$                         &$ 68.898 $  &$ 69.158^{+5.517}_{-6.563}   $   &             & $64.738 $      &$ 64.662^{+0.673}_{-0.6609}  $     &           \\       \hline
       
PI     & $\Omega_b{h}^2$                  &$ 0.0219 $  &$ 0.0220^{+0.0005}_{-0.0003} $   & $7554.198$  & $0.0211$          &  $0.0211^{+0.0001}_{-0.0001}  $   &$11624.540$ \\              [1ex]
       & $\Omega_{c} h^2 $                &$ 0.1172 $  &$ 0.1164^{+0.0041}_{-0.0045} $   &             & $0.1247$          &  $ 0.1164^{+0.0016}_{-0.0016} $   & \\                         [1ex]
       &  $\Omega_\Lambda$                &$ 0.6914 $  &$ 0.6953^{+0.0273}_{-0.0218} $   &             & $0.1040$          &  $0.6498^{+0.0109}_{-0.0110} $    &\\                          [1ex]
       &  $\tau$                          &$ 0.0825 $  &$ 0.0826^{+0.0118}_{-0.0137} $   &             & $0.6746$          &  $ 0.0571^{+0.0204}_{-0.0204} $   & \\                         [1ex]
       &  $100\theta$                     &$ 1.0389 $  &$ 1.0387^{+0.0020}_{-0.0020} $   &             & $1.0400$          &  $ 1.0400^{+0.0003}_{-0.0003} $   &  \\                        [1ex]
       &  $z_{re}$                        &$ 10.420 $  &$ 10.343^{+1.093}_{-1.083}   $   &             & $9.403$           &  $ 8.1973^{+2.4195}_{-1.9729}   $   & \\                         [1ex]
       &  $a_0$                           &$ 0.0097 $  &$ 0.0104^{+0.0041}_{-0.0066} $   &             & $0.0070$          &  $ 0.0065^{+0.0020}_{-0.0019} $   & \\                         [1ex]
       &  $\phi_0$                        &$ 1.9654 $  &$ 1.9661^{+0.0039}_{-0.0011} $   &             & $1.9628$          &  $ 1.9630^{+0.0013}_{-0.0015} $   & \\                         [1ex]
       &  $\ln[10^{10}m^2]$               &$-9.8376 $  &$ -10.137^{+0.626}_{-1.196}  $   &             & $-9.1350$         &  $-9.2039^{+0.3298}_{-0.4269}  $    & \\                         [1ex]
       &  $H_0$                           &$ 67.336 $  &$ 67.694^{+1.919}_{-1.738}   $   &             & $64.450$          &  $  64.618^{+0.6591}_{-0.7400} $  & \\       \hline            
                     
\end{tabular}
}
\caption{The best fit and mean values for power law (PL) model and Punctuated Inflation (PI) model obtained using WMAP9 year
and Planck data sets.}

\label{estimates}
\end{table*}
\begin{figure*}
\centering
\includegraphics[width=11cm,height=11cm,keepaspectratio]{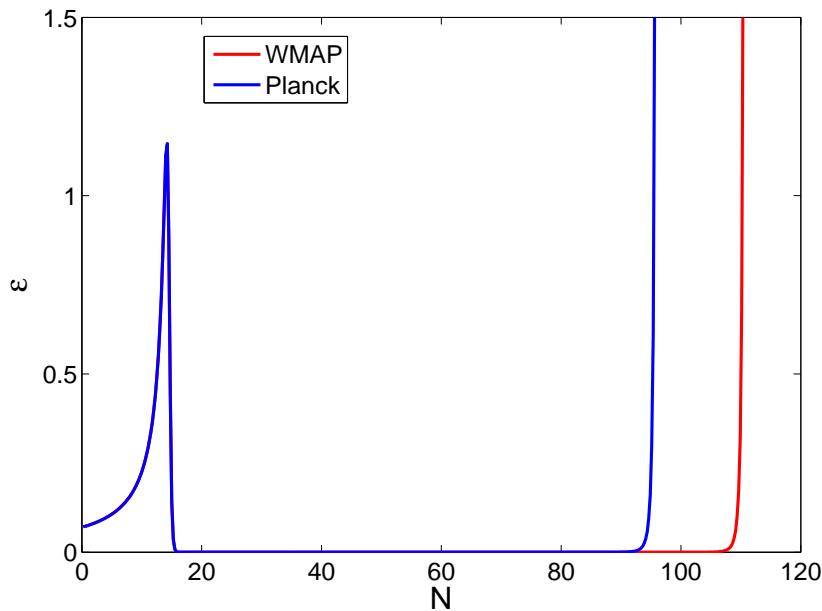}
\caption{Evolution of the slow roll parameters $\epsilon$ as a function of e-folds $N$ in case of PI model for WMAP9 and Planck data.
The slow roll approximation  is violated (i.e $\epsilon>1$) for an e-fold around $N=13$ - $14$ e-folds. It can be seen that the evolution of $\epsilon$ is identical for WMAP9 and Planck data sets except at the 
late stage of inflation. Also note that here inflation ends at relatively large value of $N$ (i.e $N=112$ and $N=92$ for WMAP9 and Planck data sets respectively).
}
\label{epsilon}
\end{figure*}

\begin{figure*}
\begin{center}
\includegraphics[width=15cm,height=16cm]{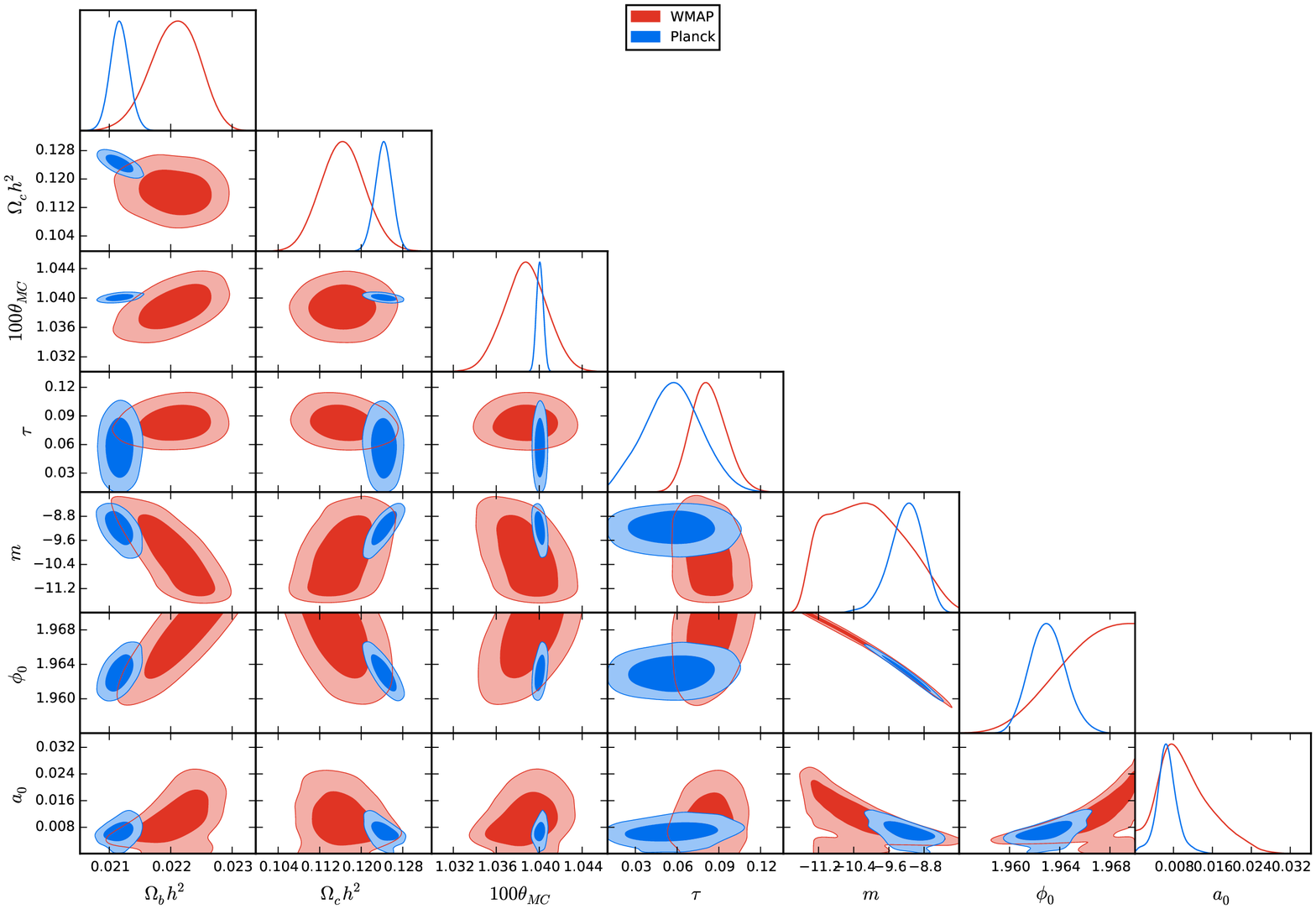}
\caption{Two dimensional joint posterior probability distributions and one dimensional
marginal posterior probability distribution  for the parameters of the punctuated inflation along with other cosmological parameters.} 
\label{posterior}
\end{center}
\end{figure*}

\section{Results and discussion}
\subsection{Best fit parameters and joint constraints}
\begin{figure*}
\centering
\includegraphics[width=11cm,height=11cm,keepaspectratio]{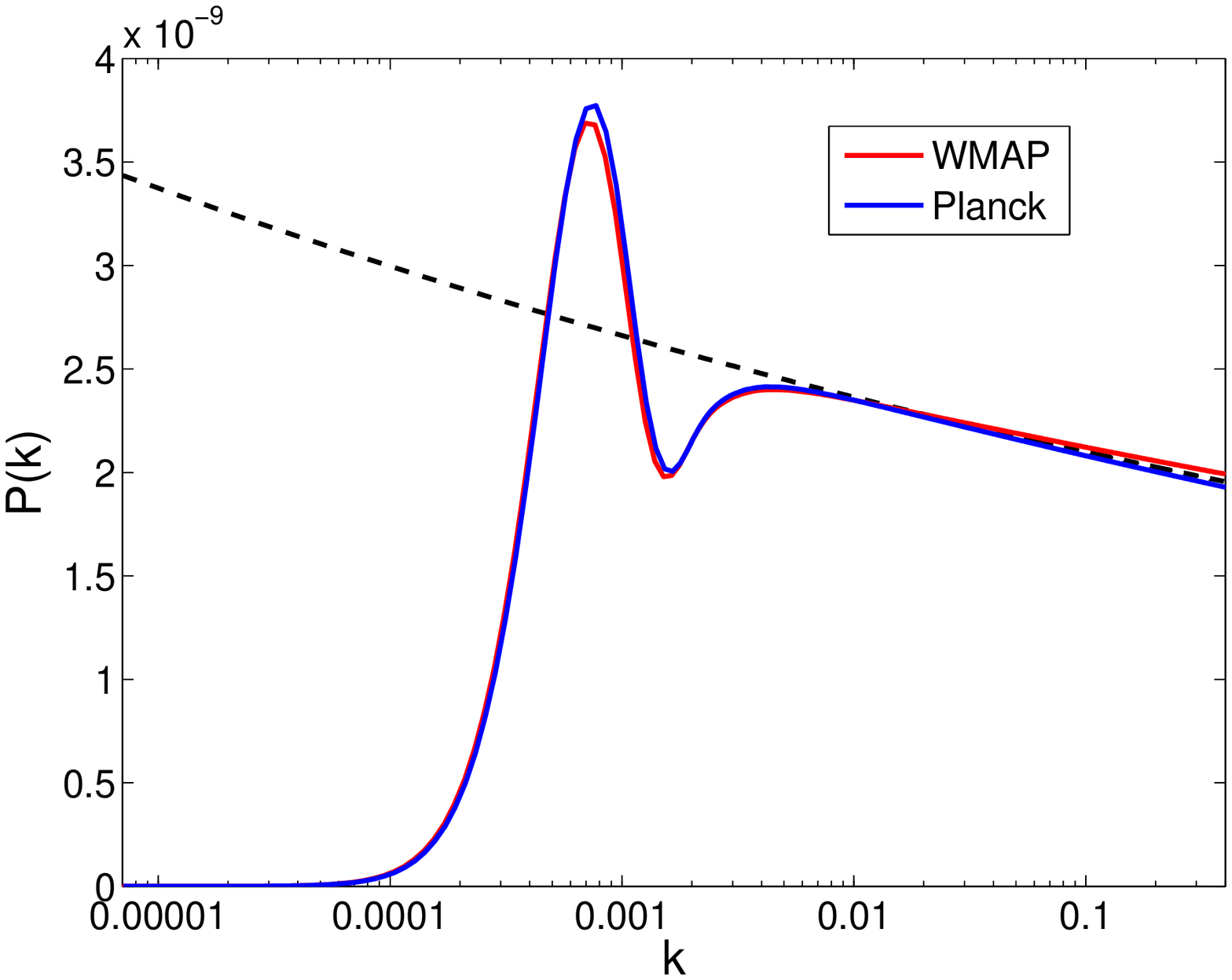}
\caption{The best fit primordial spectra for PI model using WMAP9 (red line) and Planck (blue line) data sets. 
Note for both cases, we find sharp cut off around $ k\approx4\times10^{-4}$ followed by a bump and suppression. At large $k$, PI primordial power spectrum matches with the standard power law model shown with the black dashed line.}
\label{bestfit_pk}
\end{figure*}

\begin{figure*}
\centering
\includegraphics[width=11cm,height=11cm,keepaspectratio]{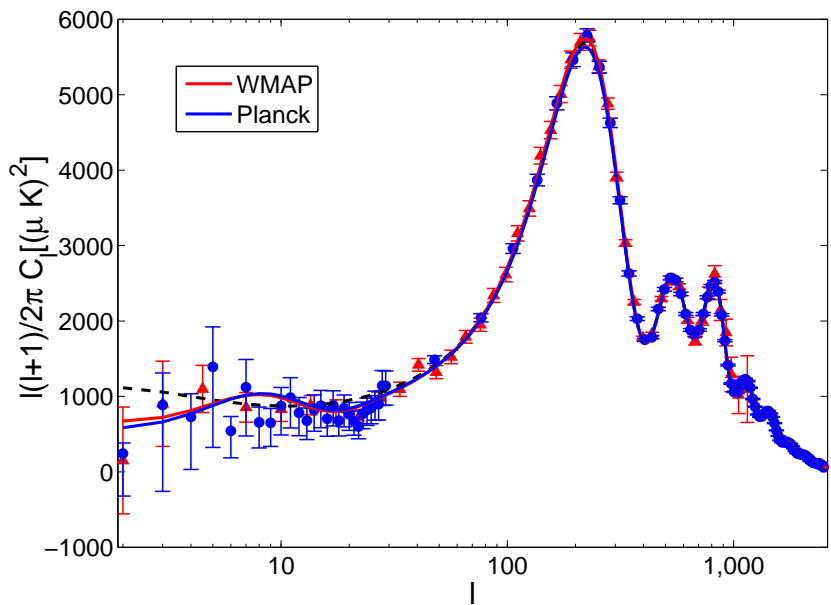}
\caption{The best fit angular power spectra $C_l^{TT}$ for PI model using WMAP9 and Planck data sets. 
For comparison, we have also plotted  $C_l^{TT}$ using power law model with black bashed line. 
The observed data points for WMAP9 and Planck data are also shown by red triangles and blue dots respectively.}
\label{bestfit_cl1}
\end{figure*}

\begin{figure*}
\centering
\includegraphics[width=11cm,height=11cm,keepaspectratio]{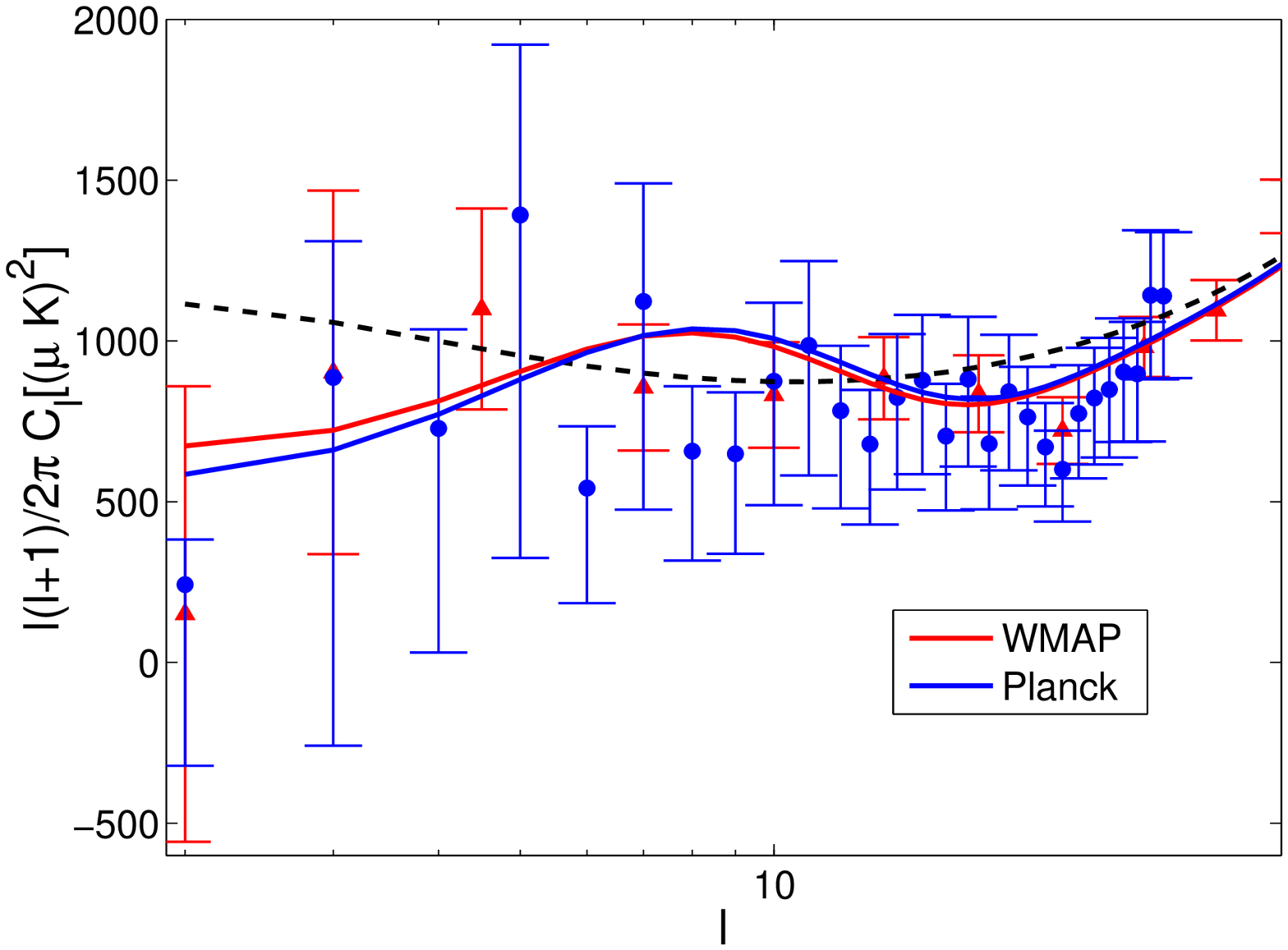}
\caption{Same as in Fig.~(\ref{bestfit_cl1}) at low-$\ell$. From this figure it is clear that the 
PI model induces power suppression up to a large values of $l\leq40$ with a bump around $5\leq \ell\leq11$ which results in the better improvement of fit in case of Planck data
compared to WMAP9 data}.
\label{bestfit_cl2}
\end{figure*}

In table~\ref{estimates}, we present the estimates of all the parameters  in terms of the best fit values and their mean values with 1-$\sigma$ error bars 
for both PL model and PI model using  WMAP9 and Planck data. 
Figure~\ref{epsilon} shows the evolution of slow roll parameter $\epsilon$ as a function of e-folds $N$ for the best fit parameters of WMAP9 and Planck data set.
As can be seen from figure~\ref{epsilon}, the slow roll approximation is violated for an e-fold around 13-14 e-folds. 
Figure~\ref{posterior} shows marginalized posterior distribution and two dimensional posterior distributions along with contours at 68\% and 95\% of the cosmological parameters for WMAP9 and Planck data set. Figure~\ref{bestfit_pk} shows best fit primordial spectra for PI model using WMAP9 and Planck data set. For comparison we have also plotted the best fit primordial power spectrum for the pure power law case. Figures~\ref{bestfit_cl1} \& \ref{bestfit_cl2} show the corresponding angular power spectra $ C_{\ell}^{TT}$ for the best fit values of PI model parameters and other standard cosmological parameters.

From table~\ref{estimates},  we see that PI model  gives better fit to the data than the standard power law model at the cost of one extra parameter. For WMAP9 the improvement is  $\Delta \chi^2\sim 3.6$ and for  Planck data set 
the  improvement in fit is $\Delta \chi^2\sim 5.4$. In the next section, we will discuss the significance of these fits using $AIC$. 
From figure~\ref{posterior}, it is clear that we are able to obtain good bounds on all the parameters and that both WMAP9 and Planck  data sets are consistent with each other. In fact, we were able to put tighter constraints on the parameters with the Planck data set. 
It follows from figure~\ref{bestfit_pk}, that the main feature of the primordial power spectrum for PI model is 
the existence of sharp cut-off around $k\approx4\times10^{-4}$ followed by a bump around $k\approx 0.001$ and suppression around $k\approx 0.002$ in both WMAP9 and Planck cases. This  corresponds to sharp cut-off  in the  $ C_{\ell}^{TT}$ spectrum 
up to multipoles $\ell \leq 4$ followed by the bump ($5\leq \ell\leq11$) and suppression ($12\leq \ell\leq40$) as shown figure~\ref{bestfit_cl1}. 

\subsection{Model comparison}
To judge whether a model is preferred by data, we use Akaike information criterion ($AIC$) which incorporates trade-off between the goodness of 
fit and additional complexity of the model \cite{Akaike1974_aic,2004MNRAS.351L..49L,2003JCAP...07..002C}:
\begin{equation}
AIC = -2 \ln {\mathcal L_{max}}({\bf d}|{\bf \theta}) + 2 N,
\label{aic_def}
\end{equation}
where $N$ is the number of free parameters and $\chi^2=-2\ln{\mathcal{L}_{max}}$, $\mathcal{L}_{max}$ being best fit likelihood value of the given model. In the above equation, first term represents the quality of the model fit and second term represent the 
model complexity. The $AIC$ is based on approximation to the Kullback-Leibler information entropy \cite{Anderson2002,2013PhRvD..88b3522G}. 
The preferred model is the one which has a minimum value of $AIC$  and $\Delta AIC \equiv AIC_{i}-AIC_{min}$ represents preference of model $i$ over the best fit model (model with minimum value of $AIC$ i.e $AIC_{min}$). 
Models with $\Delta AIC \leq 2$ have substantial support, models with $4 < \Delta AIC <7$ have considerably less support and 
those with $\Delta AIC > 10$ have essentially no support compared to best fit model \cite{Kenneth2002}. 

For Planck data set, we find $\Delta AIC=3.40$ which implies that PI model is moderately preferred compared to PL model despite having an additional parameter. However, for the WMAP9 only data set, we find that both models are equally favorable.

\section{Conclusions}
Although power law  model has emerged as the most successful model consistent with recent observations, it could not explain certain anomalies. One of these anomalies is low CMB power at large angular scales in the CMB power spectrum. Different models have been used, having their own virtues and deficiencies, to account for this power loss.  For example, in our previous work \cite{Iqbal2015}, we studied various inflationary models using different initial condition (like kinetic or radiation dominated era). 
 
Motivated by the initial condition independence, in this work, we have used the punctuated inflationary model in which a brief period of fast roll is sandwiched between two stages of slow-roll inflation. Markov Chain Monte Carlo analysis has been performed to determine posterior distributions and the values of model parameters that provide best fit to WMAP9 and Planck data for CMB angular power spectrum. We found that PI model gives better  fit to the data ($\Delta\chi^2 \approx 3.6$ for WMAP9 and $\Delta\chi^2 \approx 5.4$ for Planck) than the standard $\Lambda$CDM model with a  featureless, primordial power spectrum at the cost of one extra parameter. Further, we used Akaki information criteria for model comparison which showed that for Planck data, PI model (having an additional parameter) is moderately preferred over PL model while as for WMAP9 data set only,  both models are equally favorable.

There are models which superimpose oscillations on the  power spectrum and which could produce $\Delta\chi^2 \approx 10$ - $16$ \cite{2014PhRvD..89f3537M,2014PhRvD..89f3536M,Ade2015b}. However, in these cases the fitting is forced in the entire spectrum and the number of extra parameters involved are three or more. There also remains problem of over fitting  data with such models. Similarly, many models have also been proposed with a step like feature in the inflationary potential \cite{Starobinsky1982,2001PhRvD..64l3514A,2005JCAP...07..015G} which are also found to better fit the data. Therefore, it is clear that significant improvement in both data and modeling is required to establish supremacy of one model over the other.

\section*{Acknowledgments}
AI and MHQ  would like to thank IUCAA for its hospitality during their stay. 
AI and MHQ would also like to thank Dhiraj Kumar Hazra and Jayanti Prasad for continuous exchange of emails during initial stage of the project. We acknowledge the use of IUCAA's high performance computing facility for carrying out this work. The work of AI and MM was supported by DST Project Grant No. SR/S2/HEP-29/2012. The authors would like to thank the anonymous referee for useful comments.

\bibliographystyle{JHEP}
\bibliography{pps}
\end{document}